\documentstyle[12pt]{article}                                                            
\newlength{\dinwidth}                                                            
\newlength{\dinmargin}                                                            
\def\lapproxeq{\lower .7ex\hbox{$\;\stackrel{\textstyle                                                            
<}{\sim}\;$}}                                                            
\def\gapproxeq{\lower .7ex\hbox{$\;\stackrel{\textstyle                                                            
>}{\sim}\;$}}   
                                                                                          
\def\be{\begin{equation}}                                                            
\def\ee{\end{equation}}                                                            
\def\bea{\begin{eqnarray}}                                                            
\def\eea{\end{eqnarray}}

\begin{document}    
\titlepage                                               
\begin{center}
{\Large \bf Comment on an invisible Higgs boson and 50 GeV 
neutrino} \\

\vspace*{1cm}
V. A. Khoze \\
\vspace*{0.5cm}                                                    
Department of Physics and IPPP, University of Durham, Durham, DH1 3LE  
\end{center}

\vspace*{3cm}

%\vspace*{0.3cm}

\noindent 
It has been recently suggested \cite{MM} that the fourth generation of leptons
and quarks is not excluded by the precision $Z$ boson measurements, provided that
the mass of the fourth neutrino  is around 50 GeV and that 
its mixing with the neutrinos of the first three generations is negligibly small.

We would like to note that in this case the decay rate of the Standard Model Higgs 
boson into the pair of such neutrinos would be about two orders of magnitude higher than 
than the $H \rightarrow b\bar{b}$ rate. Thus, the predominant decay mode of the 
Higgs would be invisible, which would require a special strategy for its searches, 
see, for example, \cite{EZ}.

\bigskip

I am very grateful to V. A. Novikov, M. I. Vysotsky and L. B. Okun for discussions.


\vspace*{1cm}

\begin{thebibliography}{xx} 
\bibitem{MM} 
M.~Maltoni, V.~A.~Novikov, L.~B.~Okun, A.~N.~Rozanov and M.~I.~Vysotsky,
%``Extra quark-lepton generations and precision measurements,''
Phys.\ Lett.\ B {\bf 476} (2000) 107
[hep-ph/9911535];
%%CITATION = HEP-PH 9911535;%%

V.~A.~Ilyin, M.~Maltoni, V.~A.~Novikov, L.~B.~Okun, A.~N.~Rozanov and M.~I.~Vysotsky,
%``On the search for 50-GeV neutrinos,''
Phys.\ Lett.\ B {\bf 503} (2001) 126
[hep-ph/0006324].
%%CITATION = HEP-PH 0006324;%%
\bibitem{EZ} 
O.~J.~Eboli and D.~Zeppenfeld,
%``Observing an invisible Higgs boson,''
Phys.\ Lett.\ B {\bf 495} (2000) 147
[hep-ph/0009158].
%%CITATION = HEP-PH 0009158;%%
\end{thebibliography}
\end{document}